\documentstyle{article}
\input math_macros
\font\tenbf=cmbx10

\font\tenrm=cmr10
\font\tenit=cmti10
\font\elevenbf=cmbx10 scaled\magstep 1
\font\elevenrm=cmr10 scaled\magstep 1
\font\elevenit=cmti10 scaled\magstep 1

\font\ninerm=cmr9

\def \m{\rm MeV}
\renewenvironment{thebibliography}[1]
 { \elevenrm
   \begin{list}{\arabic{enumi}.}
    {\usecounter{enumi} \setlength{\parsep}{0pt}
     \setlength{\itemsep}{3pt} \settowidth{\labelwidth}{#1.}
     \sloppy
    }}{\end{list}}

\begin{document}
\begin{center}{{\tenbf ELECTROWEAK MEASUREMENTS AND TOP QUARK MASS
LIMITS\footnote{\ninerm Presented at DPF 92 Meeting, Fermilab, November,
1992.}\\}
\vspace{-1in}
\rightline{EFI 92-58}
\rightline{November 1992}
\bigskip
\vglue 2.0cm
{\tenrm JONATHAN L. ROSNER\\}
\baselineskip=13pt
{\tenit Enrico Fermi Institute and Department of Physics,
University of Chicago\\}
\baselineskip=12pt
{\tenit 5640 S. Ellis Ave., Chicago, IL 60637, USA\\}
\vglue 0.8cm
{\tenrm ABSTRACT}}
\end{center}
\vglue 0.3cm
{\rightskip=3pc
 \leftskip=3pc
 \tenrm\baselineskip=12pt
 \noindent
The agreement of electroweak measurements with theory places limits on the
masses of the top quark and the $W$ boson. It is shown how these limits arise
and what constraints various measurements (particularly a top quark mass
determination) would provide on the theory. The degree to which present and
future measurements can constrain the Higgs boson mass is examined.
\vglue 0.6cm}
{\elevenbf\noindent 1. Introduction}
\vglue 0.4cm
\baselineskip=14pt
\elevenrm
The unified description of weak and electromagnetic interactions has led to a
series of predictions which are in accord with all present data, including
recent measurements from $e^+ e^-$ collisions at the $Z$ mass, $p \bar p$
collisions at high energies, deep inelastic scattering of neutrinos, and
parity violation in atoms.  This agreement is so good that it constrains
higher-order effects, particularly that of the top quark mass.  In the
present paper we examine such constraints in the light of recent data,
and extend the analysis of a recent review \cite{RMPColl} to parameters
$S$ and $T$ which describe effects of new physics \cite{PT}.
%
\vglue 0.5cm
{\elevenbf\noindent 2. Electroweak theory and radiative corrections}
\vglue 0.2cm
The electroweak theory replaces the Fermi coupling constant,
$G_F = 1.16637 \pm 0.00002 \times 10^{-5}$ GeV$^{-2}$,
by combinations of dimensionless couplings and masses:
\begin{equation}
\frac{G_F}{\sqrt{2}} = \frac{g^2}{8M_W^2}~~,~~~
\frac{G_F}{\sqrt{2}} = \frac{g^2 +{g'}^2}{8M_Z^2} ~~~,
\end{equation}
where $(g,g')$ are the SU(2) and U(1) couplings. The electric charge $e$ is
related to $g$ and $g'$ by $e = g \sin \theta = g' \cos \theta$, so that
\begin{equation}
M_W^2 = \frac{\pi \alpha}{\sqrt{2} G_F \sin^2 \theta}~~,~~~
M_Z^2 = \frac{\pi \alpha}{\sqrt{2} G_F \sin^2 \theta \cos^2 \theta} ~~~.
\end{equation}
The value of $M_Z = 91.187 \pm 0.007$ GeV \cite{Rolandi} then can be used
to predict the value of $\theta$ in the lowest-order theory, leading to a
value of $M_W$.

The electromagnetic charge when probed at the scale of $M_W$ or $M_Z$ is
slightly stronger than that at long distances as a result of vacuum
polarization
effects.  The fine-structure constant, instead of being about 1/137, is about
1/128.  This modification is crucial in obtaining a value of $M_W$ from the
above procedure which is close to the experimental average \cite{CDF91,UA291}.

The major effect of a top quark mass is a modification of the relation
between $G_F$ and $M_Z$:
\begin{equation}
\frac{G_F}{\sqrt{2}} \rho = \frac{g^2 + {g'}^2}{8M_Z^2} ~~,~~~
\rho \simeq 1 + \frac{3G_F m_t^2}{8 \pi^2 \sqrt{2}} ~~~.
\end{equation}
The quadratic dependence on $m_t$ comes from the top quark's contribution to
$W$ and $Z$ self-energy diagrams.  No such quadratic dependence appears in
the photon vacuum polarization because of electromagnetic gauge invariance.
The relation for $M_Z$ in terms of $\theta$ now becomes
\begin{equation}
M_Z^2 = \frac{\pi \alpha}{\sqrt{2} G_F \rho \sin^2 \theta \cos^2
\theta}~~~.
\end{equation}

Higgs boson contributions to $W$ and $Z$ self-energies lead to an additional
term
\begin{equation}
\Delta \rho = - \frac{3}{8 \pi \cos^2 \theta} \ln \frac{M_H}{M_W}
\end{equation}
in $\rho$.  Now, $\theta$, $M_W$, and other electroweak observables depend on
both $m_t$ and $M_H$.  This dependence, along with present bounds on $M_W$
\cite{CDF91,UA291}, leads for $m_H < 1$ TeV to a crude upper limit of $m_t \le
200$ GeV.  The lower bound on $m_t$ (95\% c.l.) is 91 GeV \cite{CDF92}.
A measurement of $m_t$ to $\pm 5$ GeV and $m_W$ to $\pm 50$ MeV will begin
to distinguish among predictions for various Higgs masses.

Additional terms logarithmic in $m_t$ lead to modifications of the relations
written previously:
\begin{equation}
\frac{G_F}{\sqrt{2}} = (1 + \Delta Z_W ) \frac{g^2}{8 M_W^2 }~~,~~~
\frac{G_F \rho}{\sqrt{2}} = (1 + \Delta Z_Z) \frac{g^2 + {g'}^2}{8M_Z^2} ~~~,
\end{equation}
where $\Delta Z_W$ and $\Delta Z_Z$ represent the effects of variation with
momentum transfer between $q^2 = 0$ (where $G_F$ is measured) and the $W$ and
$Z$ poles (where coupling constants and masses are defined).  They may be
expressed in terms of quantities of order 1:
\begin{equation}
\Delta Z_W =
\frac{\alpha S_W}{4 \sin^2 \theta} ~;
{}~~~ \Delta Z_Z =
\frac{\alpha S_Z}{4 \sin^2 \theta \cos^2 \theta } ~~~,
\end{equation}
Similarly, $\rho = 1 + \alpha T$ can be expressed in terms of a parameter $T$
of order 1.

If one expands around nominal values of $m_t$ and $M_H$, one finds \cite{KL}
\begin{equation}
T \simeq \frac{3}{16 \pi \sin^2 \theta} \left [ \frac{m_t^2 - (140 ~{\rm GeV}
)^2}{M_W^2} \right ] - \frac{3}{8 \pi \cos^2 \theta} \ln
\frac{M_H}{100 ~{\rm GeV} } ~~~,
\end{equation}
\begin{equation}
S_W = \frac{1}{6 \pi} \left [ \ln \frac{M_H}{100~{\rm GeV} } - 2 \ln
\frac{m_t}{140~{\rm GeV} } \right ] ~~,~~~
S_Z = \frac{1}{6 \pi} \left [ \ln \frac{M_H}{100~{\rm GeV} } + 4 \ln
\frac{m_t}{140~{\rm GeV} } \right ] ~~~.
\end{equation}

\vglue 0.5cm
{\elevenbf \noindent 3. Electroweak observables and fits}
\vglue 0.2cm
The precise value of the $Z^0$ mass entails a value of
$\sin^2 \theta \equiv x_0 = 0.2323 \pm 0.0002 \pm 0.0005$
for $m_t = 140$ GeV, $M_H = 100$ GeV.  We expand a set
of electroweak obervables about this value; details are to be found in
Ref.~\cite{RMPColl}.  For example, we have \cite{MR}
\begin{equation}
\sin^2 \theta - x_0 = \frac{\alpha}{1-2 x_0} \left [ \frac{1}{4} S_Z - x_0 (1 -
x_0) T \right ] = (3.65 \times 10^{-3}) S_Z - (2.61 \times 10^{-3}) T~~~,
\end{equation}
and corresponding other expressions for $Z$ partial widths, neutral-current
to charged-current ratios in deep inelastic neutrino scattering, and weak
charges as measured in atomic parity violation.  The data are summarized
in Table I.

\renewcommand{\thetable}{\Roman{table}}
\begin{table}
\caption{Electroweak observables incorporated into a fit to the standard
electroweak theory.}
\begin{center}
\small{
\begin{tabular}{| l c c c c |} \hline
Quantity & Reference & Experimental & Nominal       & Expt. $\div$ \\
         &           & Value        & Theory$^{a)}$ & theory       \\
\hline \hline
$Q_W$ (Cs) & b) & $-71.04 \pm 1.81$& $-73.20$ & $0.970\pm 0.025$ \\
$M_W ({\rm GeV})$& c) & $80.14\pm 0.27$ & $80.21^{b)}$& $ 0.999 \pm 0.003$ \\
$N_\nu$ [from $\Gamma (Z \to \nu \bar \nu )$] & d) & $3.04 \pm 0.04$& $3$ &
$1.013 \pm 0.013 $ \\
$\Gamma (Z \to l^+l^- ) (\m )$& e) & $83.52 \pm 0.33$& $83.6$& $0.999 \pm
0.004 $\\
$\Gamma (Z \to ~{\rm all})(\m )$ & d) & $2492\pm 7$ & $2488\pm 6$&
$1.001\pm 0.004 $ \\
$\bar x$ (asymms., $\tau$ pol.)& d) & $0.2324\pm 0.0011$& $0.2322$& $1.001 \pm
0.005$ \\
$\bar x (q \bar q$ asymm.)& d) & $0.2323\pm 0.0032$ & $0.2322 $ & $1.000 \pm
0.014$ \\
$\bar x (\vec e D)$ & b) & $0.224\pm 0.020$ & $0.2322$ & $0.965\pm 0.086$\\
$\bar x (\vec e C)$& b) & $0.20\pm 0.05$ & $0.2322$ & $0.86 \pm 0.22$ \\
$\bar x \left [ \sigma (\nu_\mu^{(-)}) \right]$ & f) &$0.232\pm 0.009$ &
$0.2322$ & $1.00 \pm 0.04$ \\
$M_W$ (MeV) from $R_\nu$ & g) & $80.32 \pm 0.32$ & $80.21$ & $1.001 \pm
0.004$ \\
$R_{\bar \nu}$& b) & $0.387 \pm 0.009$ & $0.376$ & $1.02 \pm 0.02$ \\
$\bar x~(A_{LR}$ at SLC) & h) & $0.2378 \pm 0.0056$ & 0.2322 & $1.024 \pm
0.024$ \\
\hline
\end{tabular}
}
\end{center}
\bigskip
\small{
\leftline{$^{a)}$For $m_t = 140 ~{\rm GeV}, ~ M_H = 100 ~{\rm GeV}$.}
\leftline{$^{b)}$As in Ref.~\cite{RMPColl}.}
\leftline{$^{c)}$Raised from value in Ref.~\cite{RMPColl} as a result of new
$M_Z$ measurement \cite{Rolandi}.}
\leftline{$^{d)}$Ref.~\cite{Rolandi}.}
\leftline{$^{e)}$F. Merritt, Seminar, Univ. of Chicago, April, 1992.}
\leftline{$^{f)}$New CHARM II value: Ref.~\cite{Rolandi} and G. R\"adel,
this conference.}
\leftline{$^{g)}$Based on CCFR value of $1 - (M_W/M_Z)^2 = 0.2242 \pm
0.0057$ \cite{Rolandi}.}
\leftline{$^{h)}$C. Baltay, this conference.  Value not included in fit.}
}
\end{table}

Based on the data in Table I, we obtain $\chi^2$ for specific values of
$M_H$ as a function of $m_t$.  The results are shown in Fig.~1.
The minimum $\chi^2$ values for $M_H = (100,~
300,~1000)$ GeV are (4.34, 4.33, 4.33), corresponding to $m_t = (144 \pm 16,~
160 \pm 15,~177 \pm 14)$ GeV.  The lack of
preference for any particular Higgs boson mass stands in
contrast to other fits \cite{Schaile,Harton} in which a slight
(but not significant) tilt in favor of
low Higgs mass occurs.  This tilt has been
traced to slightly different input values for the forward-backward asymmetry
for $b$ quark production and for the leptonic width of the $Z$.

A fit based on the degrees of freedom $S$ and $T$ of Ref.~\cite{PT} was also
performed.  (Here we have assumed $S = S_W = S_Z$, as occurs when one has
extra degenerate doublets.)  The results are shown in Fig.~2.

The elongated nature of the ellipses illustrates the absence of any preference
for a specific Higgs mass.  One can change the Higgs mass without much penalty
as long as the top
quark mass changes in a compensating way.  For $M_H < 1$ TeV (an approximate
upper bound resulting from unitarity), we see from Fig.~2 that one can still
only conclude $m_t < 200$ GeV, but with 90\% confidence.

\begin{figure}
\vspace{2.75in}
\caption{Values of $\chi^2$ for fit to 12 electroweak observables.  From left
to right, the curves correspond to $M_H = 100,~300$, and 1000 GeV.}
\end{figure}

\begin{figure}
\vspace{2.75in}
\caption{Fit to parameters $S$ and $T$ based on data of Table I.  Plotted point
corresponds to minimum $\chi^2$; inner and outer ellipses correspond to 68\%
and 90\% c.l. limits. Standard model curves correspond, from left to right, to
$M_H = 100,~300$, and 1000 GeV.  Ticks on these curves, from bottom to top,
correspond to $m_t = 100,~140,~180,~220$, and 260 GeV.}
\end{figure}

\vglue 0.5cm
{\elevenbf \noindent 4. Conclusions \hfil}
\vglue 0.4cm
We have shown that the top quark mass is limited by today's electroweak data to
be less than about 200 GeV.  Stronger limits are to be mistrusted. A plot in
$S$ and $T$ shows no particular preference for any sign of $S$ but implies $S
<1$ at the 90\% confidence level.  The discovery of the top quark and the
measurement of its mass remain the highest priority for obtaining further
information about the electroweak theory.
\vglue 0.5cm
{\elevenbf \noindent 5. Acknowledgements \hfil}
\vglue 0.4cm
I thank Ugo Fano, Henry Frisch, John Harton, Chris Hill, John Huth, Bill
Marciano, Jim Pilcher, Dorothea Schaile, and Jack Steinberger for
helpful discussions. This work was supported in part by the United States
Department of Energy under grant No. DE AC02 90ER40560.
\vglue 0.5cm
{\elevenbf\noindent 6. References \hfil}

\end{document}